\documentclass[]{emulateapj}
\usepackage{latexsym,natbib}
\usepackage{graphicx}

\begin{document}
\title{Possible twin \MakeTextLowercase{k}H\MakeTextLowercase{z} QUASI-PERIODIC OSCILLATIONS IN THE ACCRETING MILLISECOND X-RAY PULSAR IGR J17511--3057}
\author{Maithili Kalamkar,  Diego Altamirano,  and M. van der Klis} 
\affiliation{Astronomical Institute, ``Anton Pannekoek'', University of Amsterdam, Science Park 904, 1098 XH, Amsterdam, The Netherlands}
\email{m.n.kalamkar@uva.nl}
\date{ Received 2010 June 4; Accepted 2010 December 8}
\begin{abstract}
We report on the aperiodic X-ray timing and color behavior of the accreting millisecond X-ray pulsar (AMXP) IGR J17511--3057, using all the pointed observations obtained with the {\it{Rossi X-ray Timing Explorer}} Proportional Counter Array since the source's discovery on 2009 September 12. The source can be classified as an atoll source on the basis of the color and timing characteristics. It was in the hard state during the entire outburst. In the beginning and at the end of the outburst, the source exhibited what appear to be twin kHz quasi periodic oscillations (QPOs). The separation $\Delta\nu$ between the twin QPOs is $\sim$ 120 Hz. Contrary to expectations for slow rotators, instead of being close to the 244.8 Hz spin frequency, it is close to half the spin frequency. However, identification of the QPOs is not certain as the source does not fit perfectly in the existing scheme of correlations of aperiodic variability frequencies seen in neutron star low mass X-ray binaries (NS LMXBs), nor can a single shift factor make it fit as has been reported for other AMXPs. These results indicate that IGR J17511-3057 is a unique source differing from other AMXPs and could play a key role in advancing our understanding of not only AMXPs, but also NS LMXBs in general.     
\end{abstract} 
\section{Introduction}
\label{sec-intro}
Low-mass X-ray binaries (LMXBs) are neutron star (NS) or black hole systems with low-mass ({\it{M}} $\leq$ 1$M_\odot$) companion stars. Out of the nearly 200 LMXBs known so far \citep{Liu2007}, 13 are accreting millisecond X-ray pulsars (AMXPs), i.e., they have shown coherent millisecond pulsations \citep{Patruno2010}. The neutron stars in LMXBs are believed to be spun up by accretion to millisecond periods \citep[see, e.g.,][]{Bhattacharya1991}, but why only some LMXBs appear as AMXPs is still an open question. These systems can be studied through the spectral (color-color diagram - CD) and timing (Fourier analysis) properties of their X-ray emission. Based on the paths traced on the CD and associated variability, NS LMXBs are classified either as {\it{Z}} or atoll source \citep{Hasinger1989}. Correlated with the position of the source in the CD, the Fourier power spectra of the X-ray flux variations exhibit different variability components. Apart from coherent pulsations, the power spectra also exhibit aperiodic phenomena: broad components (noise components) and narrow components (quasi periodic oscillations; QPOs) \citep[see, e.g.,][for a review]{2006}. \\
\\
The coherent pulsations at the NS spin frequency are thought to be due to hot spots formed by magnetically channeled accreted matter \citep[see, e.g.,][]{Pringle1972}. The origin of the aperiodic phenomena is poorly understood. They are presumably mostly associated with inhomogeneities in the matter moving in Keplerian orbits in the accretion disk, or in the boundary layer. The timescale for matter orbiting in the strong gravity region very near to the compact object is of the order of milliseconds (the dynamical timescale $\tau = (r^3/GM)^{1/2} \sim$ 0.1 ms at distance {\it{r}} = 10 km from a compact object of mass {\it{M}} = $1.4M_\odot$). This strong gravity region can therefore be probed by the QPOs with millisecond timescales \citep[see, e.g.,][]{2006}. The first millisecond phenomena were found with {\it{Rossi X-ray Timing Explorer}} (RXTE): in Sco X-1, twin QPOs in the kHz range \citep{vdk1996} and in 4U 1728-34 similar QPOs and also burst oscillations
(oscillations near the spin frequency $\nu_s$ that occur during type I X-ray bursts; see, e.g., \citeauthor{Strohmayer2006} 2006 for a review). In many LMXBs, $\Delta\nu$, the difference between the twin kHz QPO frequencies, was at the frequency of burst oscillations $\nu_{burst}$. A beat mechanism and $\nu_{burst}$ = $\nu_s$ were suggested by \cite{Strohmayer1996b}. This led to the sonic-point beat-frequency model \citep*{Miller1998}. \\
\\
Observations of variable $\Delta\nu$ in Sco X-1 \citep{vdk1996} and later in other sources were inconsistent with the sonic-point beat-frequency model and led to a modified version \citep{Lamb2001}. Also, the relativistic precession model \citep{Stella1999} was proposed, in which $\nu_s$ plays no direct role in the formation mechanism of QPOs. Observations of LMXBs like 4U1636--53 \citep{1996a} which exhibited $\Delta\nu$ $\sim$ $\nu_{burst}$/2, clinched by the discovery of kHz QPOs in SAX J1808.8--3654 with $\Delta\nu$ $\sim$ $\nu_s$/2 by \cite{Wijnands2003}, led to the proposal of new models, involving resonances.  The relativistic resonance model \citep{Kluzniak} and spin-resonance model \citep{Lamb2003} were proposed which allowed $\Delta\nu$ = $\nu_s$ and/or $\Delta\nu$ = $\nu_s$/2. For historical accounts see \citeauthor{2006} (2006, 2008) and \citet{M'endez2007}. A model which can explain all the observations is still awaited. \\
\\
IGR J17511--3057 was discovered on 2009 September 12 during galactic bulge monitoring by INTEGRAL \citep{Baldovin2009}. It showed X-ray pulsations at 244.8 Hz during {\it{RXTE}} Proportional Counter Array (PCA) pointed observations which established its nature as an AMP \citep{Markwardt2009}. It also exhibited type I X-ray bursts and burst oscillations (\citeauthor{Watts2009a} 2009a, \citeauthor{Altamirano2010} 2010a). A minimum companion mass estimate (assuming the NS mass = 1.4$M_\odot$ and orbital inclination = $90^\circ$) is 0.13$M_\odot$ \citep{Markwardt2009} and the upper limit to the distance is 6.9 kpc \citep{Altamirano2010}. In this paper, we discuss the aperiodic variability of the AMP IGR J17511--3057. \\
\section{Observations and data analysis}
\label{sec-obdatana}
We analyzed all the 71 pointed observations with the {\it{RXTE}} PCA of IGR J17511--3057 taken between 2009 September 12 and October 6. Each observation covers one to several satellite orbits and contains up to 15 ksec of useful data for a total of $\sim$ 500 ksec. We exclude data in which the elevation (angle between the Earth's limb and the source) is less than 10$^{\circ}$ or the pointing offset (angle between source position and pointing of satellite) exceeds 0$^{\circ}$.02 \citep {Jahoda2006}. Type I X-ray bursts have been excluded from our analysis (excluding all data between 100 s before the rise and 200 s after the rise of the burst). 
\subsection{Color Analysis}
\label{sec-colorana}
The intensity and colors are obtained from  standard-2 mode data which has 16 s time resolution and 129 energy channels. The hard and soft colors are defined as the ratio of the count rates in the 9.7--16.0 keV/6.0--9.7 keV and 3.5--6.0 keV/2.0--3.5 keV bands, respectively. The intensity is the count rate in the 2.0--16.0 keV energy range. To estimate the count rates in these exact bands for each of the five proportional counter units (PCUs) of PCA, the count rates are linearly interpolated between the corresponding energy channels. Then, the background subtraction is applied to each band using the latest standard background model \footnote{See $http://heasarc.gsfc.nasa.gov/docs/xte/pcanews.html$ for more details}. The source intensity and colors are obtained at 16 s time intervals separately for each PCU. These intensity and color values are divided by the corresponding average Crab values calculated using the same method as described above, but obtained for the day closest in time to each IGR J17511--3057 observation. These Crab normalized values are then averaged over all PCUs and over time to obtain one mean value per observation. This method of normalization with Crab \citep{Kuulkers1994} is based on the assumption that Crab is constant in intensity and colors and is used to correct for differences between PCUs and gain drifts in time. 
\begin{figure}
\center
\includegraphics[width=9cm, height=10cm]{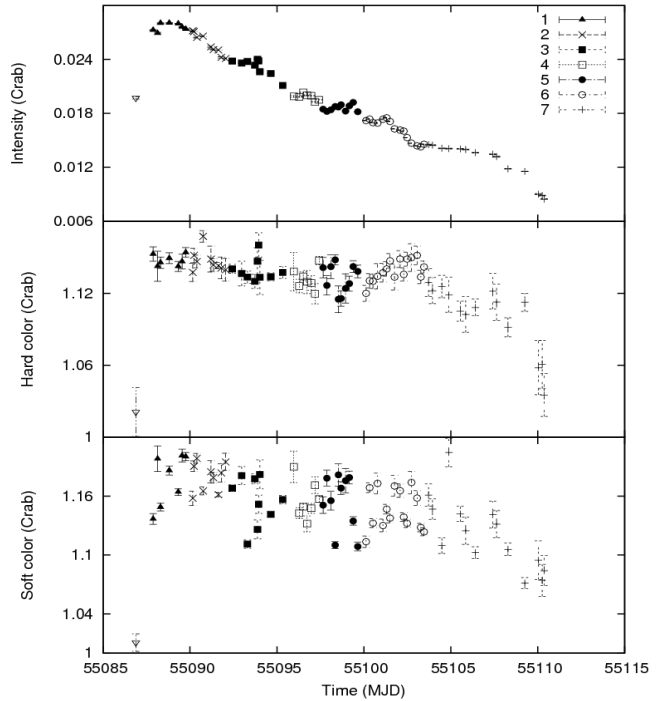}
\caption{Intensity, hard and soft colors as a function of time. Each point corresponds to one observation and different symbols identify different groups as indicated. The inverted triangle refers to the first observation in the early rise of the outburst. } \label{fig:lc}
\end{figure}
\subsection{Timing Analysis}
\label{sec-timing}
The Fourier timing analysis is done with the $\sim$ 125 $\mu$s time resolution (Nyquist frequency of 4096 Hz) event mode data in the 2--16 keV energy range. The light curve of each observation is divided into continuous segments of 128 s, which gives a lowest frequency of  1/128 s $\sim$ 8$\times 10^{-3}$ Hz. The power spectrum of each segment is constructed using the Fast Fourier Transform. These power spectra are Leahy normalized and averaged to get one power spectrum per observation. No background subtraction or dead time correction is done prior to this calculation. The Poisson noise spectrum is estimated based on the analytical function of \cite{Zhang1995} and subtracted from this average power spectrum. This power spectrum is then expressed in source fractional rms normalization \citep[see, e.g.,][]{2006} using the average background rate in the 2--16 keV band for the same time intervals.\\ 
\\
To improve statistics, we combine our observations into seven groups. The observations in a group are chosen to be close in time and have similar colors (see Section 3.1). The pulsar spike at 244.8 Hz is removed from each average power spectrum by removing all the data points in the 244--246 Hz frequency bins at full frequency resolution. The power spectrum of each group is fitted using a multi-Lorentzian function which is a sum of several Lorentzians in the '$\nu_{max}$' representation as introduced by \cite*{Belloni2002}. In this representation, the characteristic frequency is $\nu_{max} =\nu_{0}\sqrt{1+1/(4Q^{2})}$ where $\nu_{0}$ is the centroid frequency of the Lorentzian, and {\it{Q}} is the quality factor given as {\it{Q}} = $\nu_{0}$/FWHM where FWHM is the full width at half maximum. The Lorentzians are named $L_{i}$ with characteristic frequency $\nu_{i}$, where the subscript 'i' is used to identify the type of component. The components are : $L_{b}$ - break frequency, $L_{h}$ - hump, $L_{LF}$ - Low Frequency QPO, $L_{hHz}$ - hecto Hz QPO, $L_{\ell ow}$ - (perhaps) a low frequency manifestation of lower kHz QPO, $L_{\ell}$ - lower kHz QPO and $L_{u}$ - upper kHz QPO \citep*{Belloni2002}.  
\section{Results}
\subsection{Colors and Intensity}
\label{sec-resultscolor}
\begin{figure}
\includegraphics[width=8cm, height=7cm]{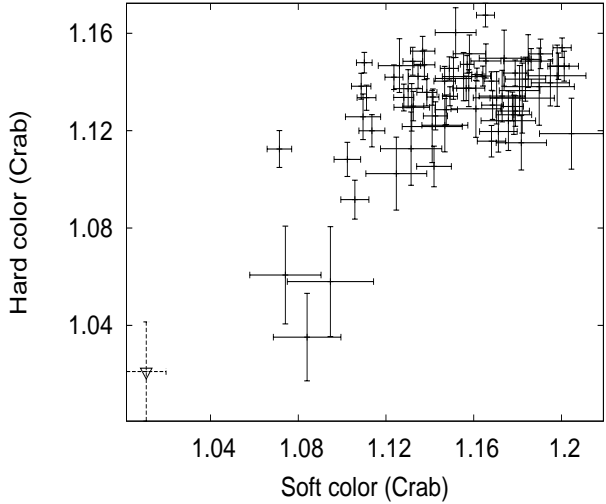}
\caption{Color-color diagram: each point corresponds to one observation. The inverted triangle refers to the first observation in the early rise of the outburst.}\label{fig:ccd}
\end{figure}

Figure  \ref{fig:lc} shows the evolution of intensity and colors with time. The seven groups are indicated in the figure. The source was first detected on 2009 September 12 and reached its peak intensity on 2009 September 14. As the intensity decreases, both the hard and soft colors decrease with time. A similar correlated behavior has been observed previously in two other AMXPs, XTE J1807--294 and XTE J1751--305 (\citeauthor{Linares2005} 2005 and \citeauthor{2005} 2005, respectively) out of the thirteen known so far. In other AMXPs and many non-pulsating transient atoll sources, the hard color remains constant while the soft color changes in correlation with intensity \citep[see, e.g.,][]{2005}. The decrease in the colors has a steeper slope after MJD $\sim$ 55103. The data in this steep fall are represented in the CD in Figure \ref{fig:ccd} by the points with a value $<1.12$ in the hard and soft colors. The light curve shows bumps at MJD $\sim$ 55090, 55093, 55098, 55102 and 55107 which are approximately traced by the hard color but not so closely by the soft color. The soft color shows considerable scatter during the entire outburst which obscures any short-term correlation with intensity. Large scatter in colors is also seen in other AMXPs, but only in the tails of their outbursts (\citeauthor{Linares2005} 2005; \citeauthor{2005} 2005). 
\subsection{Aperiodic variability}
\label{sec-resultsap}
\begin{figure*}
\includegraphics[width=6.0cm,height=6cm]{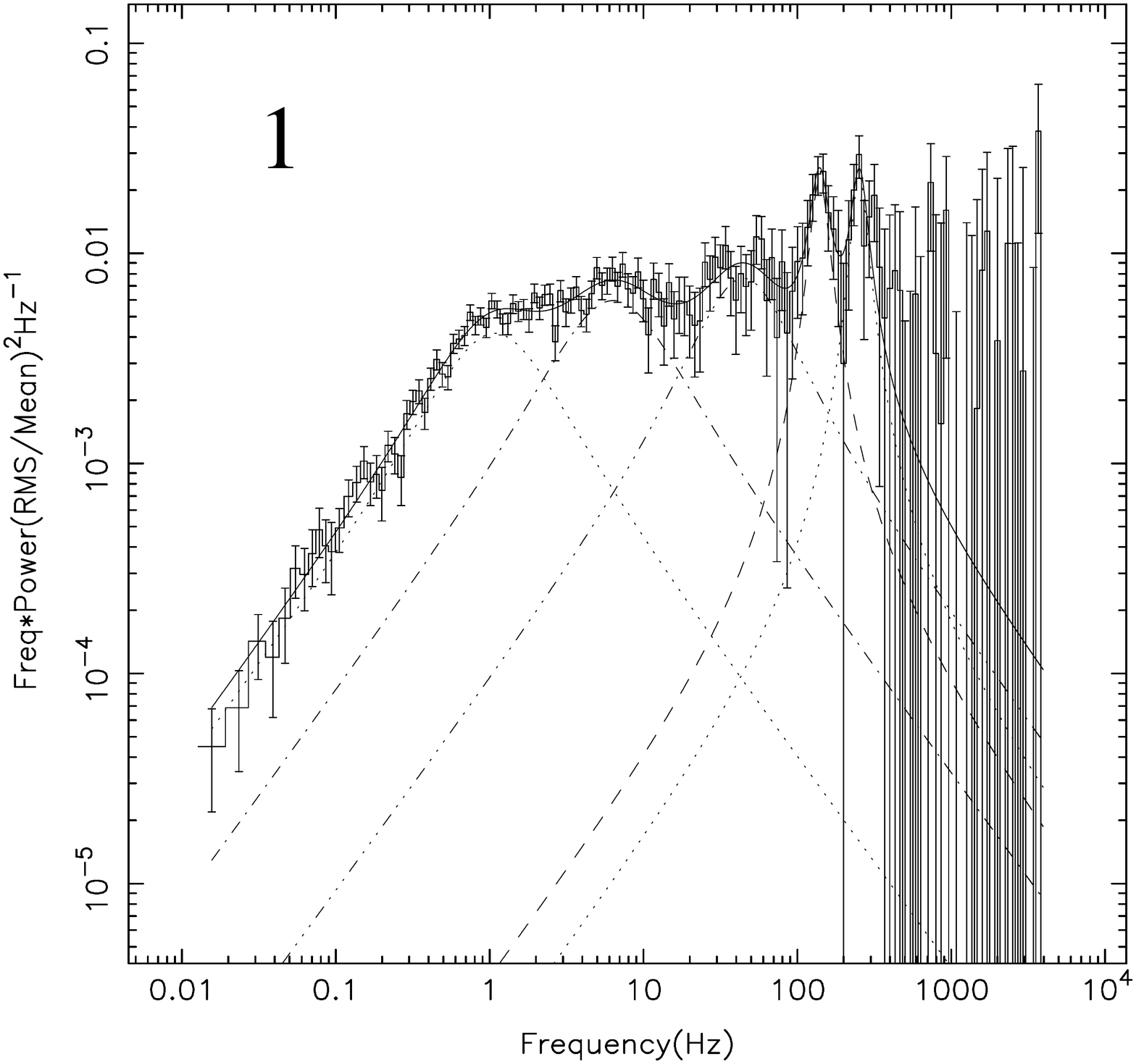}\includegraphics[width=6.0cm,height=6cm]{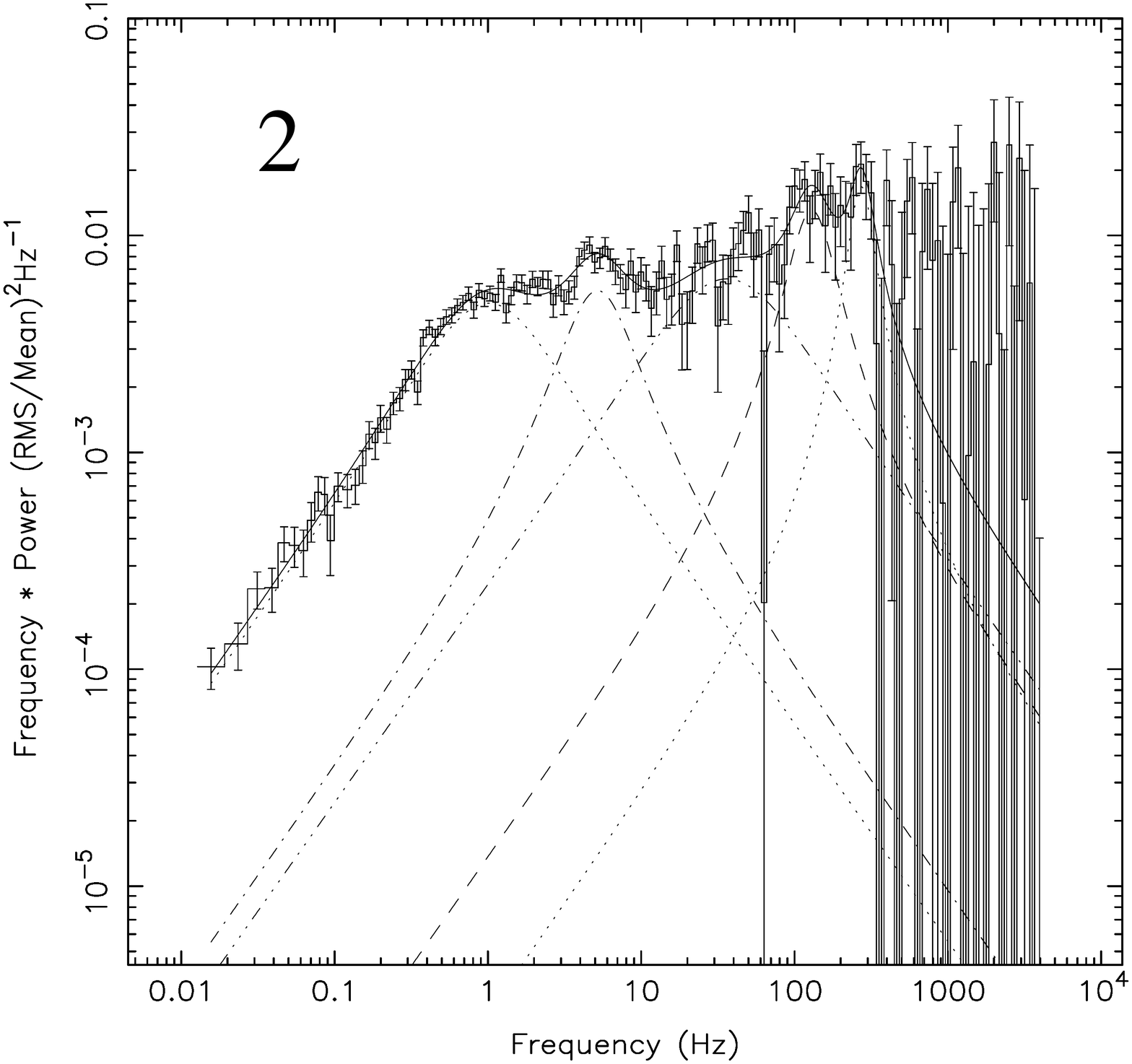}\includegraphics[width=6.0cm,height=6cm]{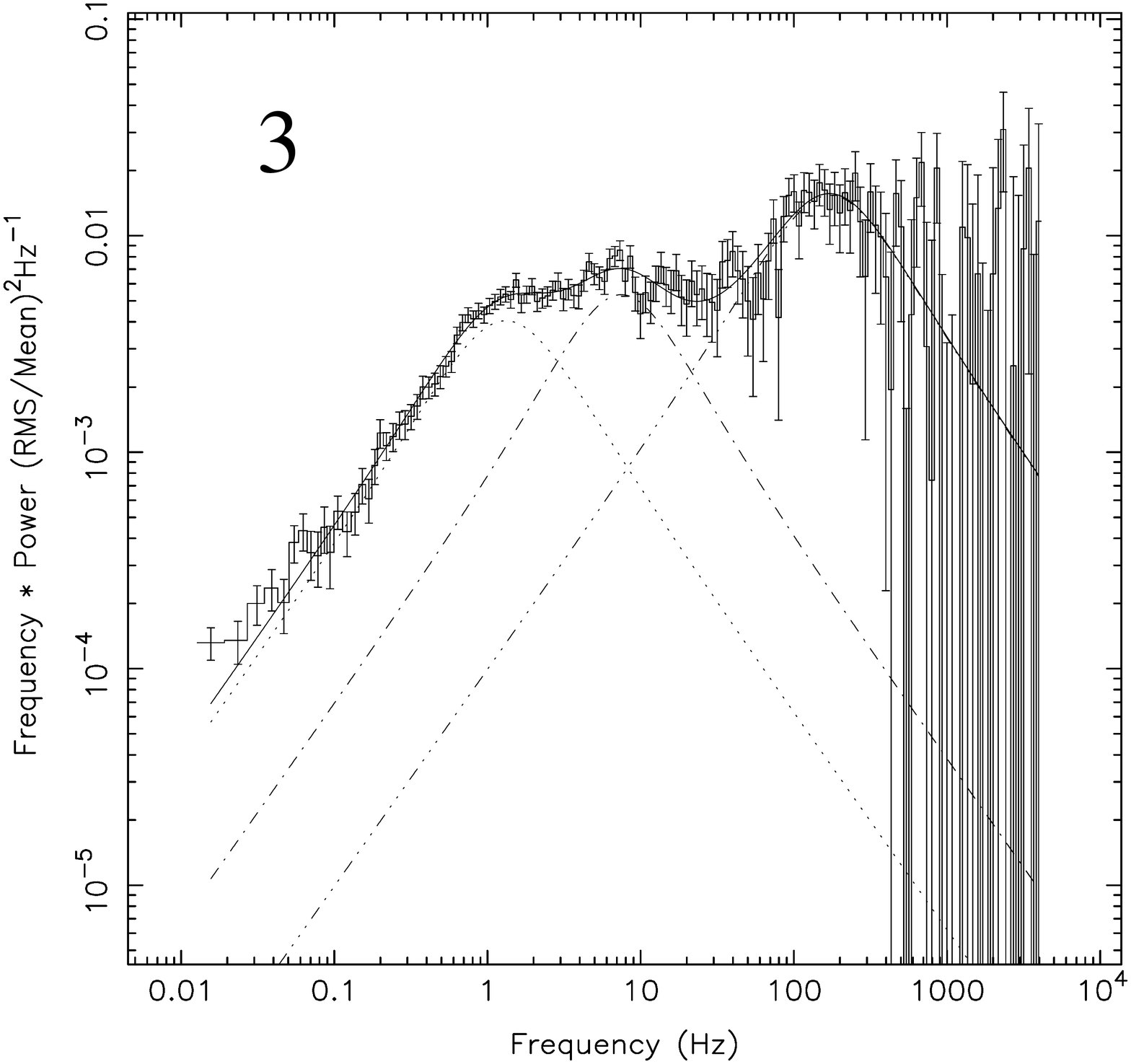}
\includegraphics[width=6.0cm,height=6cm]{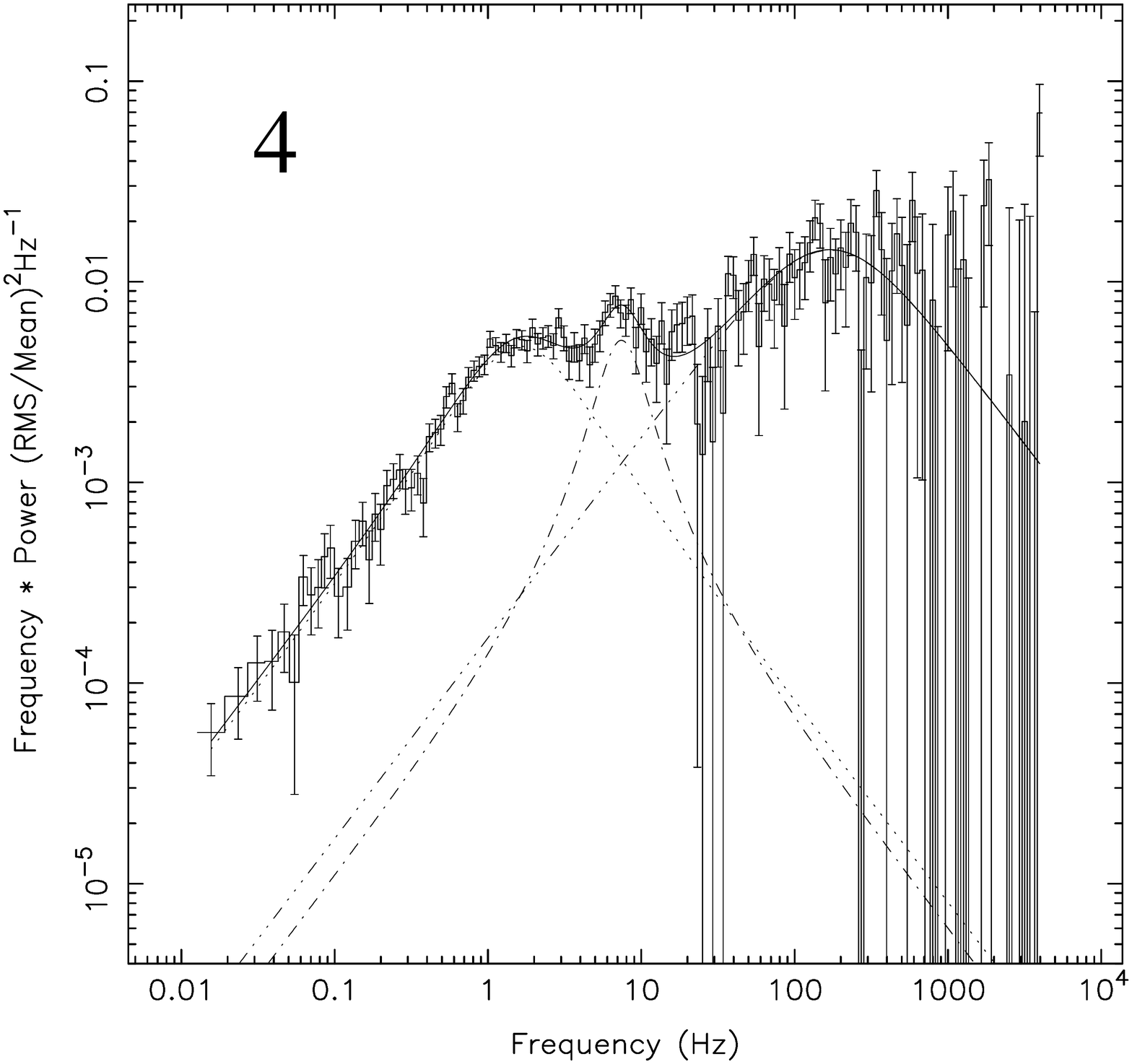}\includegraphics[width=6.0cm,height=6cm]{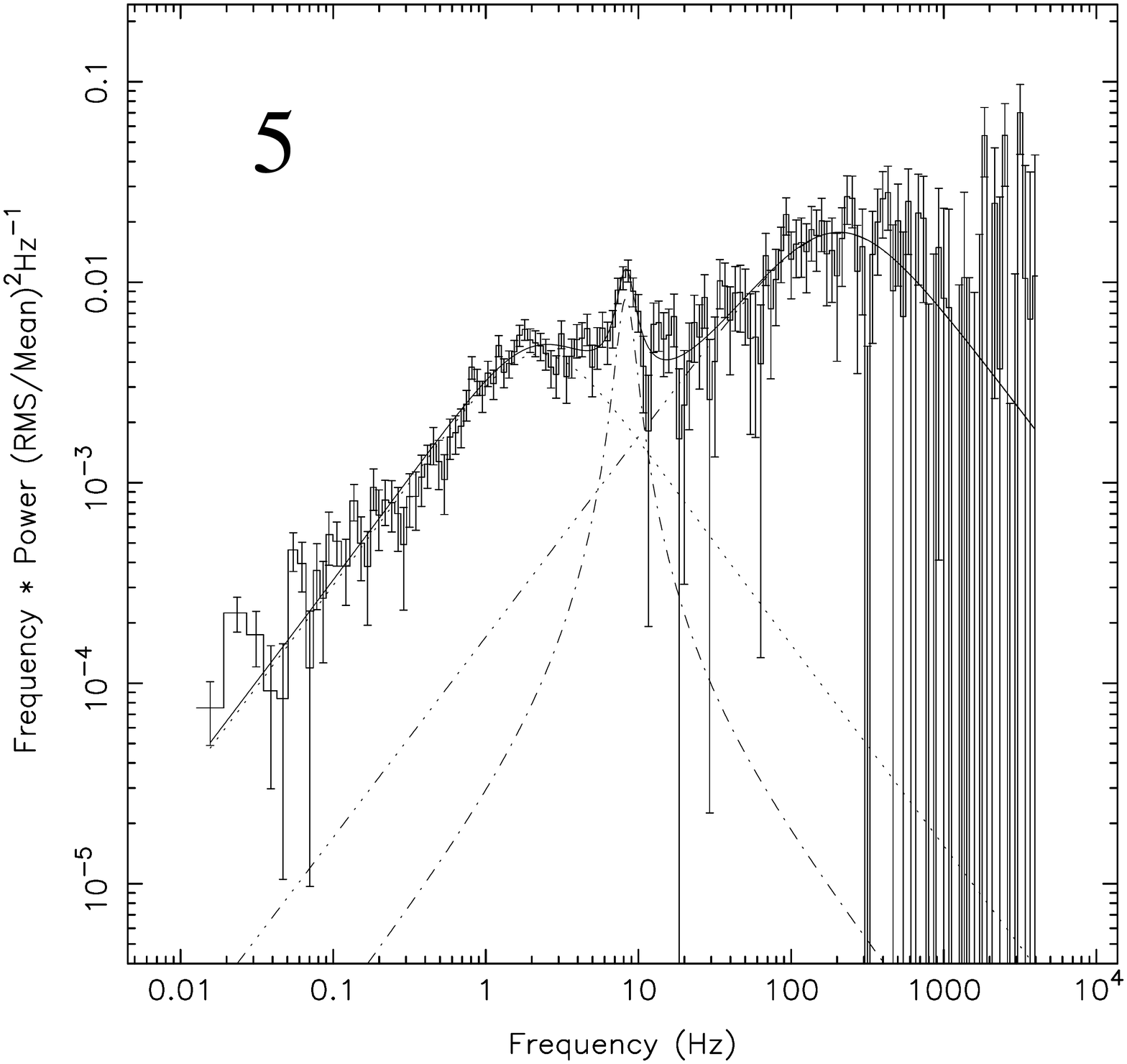}\includegraphics[width=6.0cm,height=6cm]{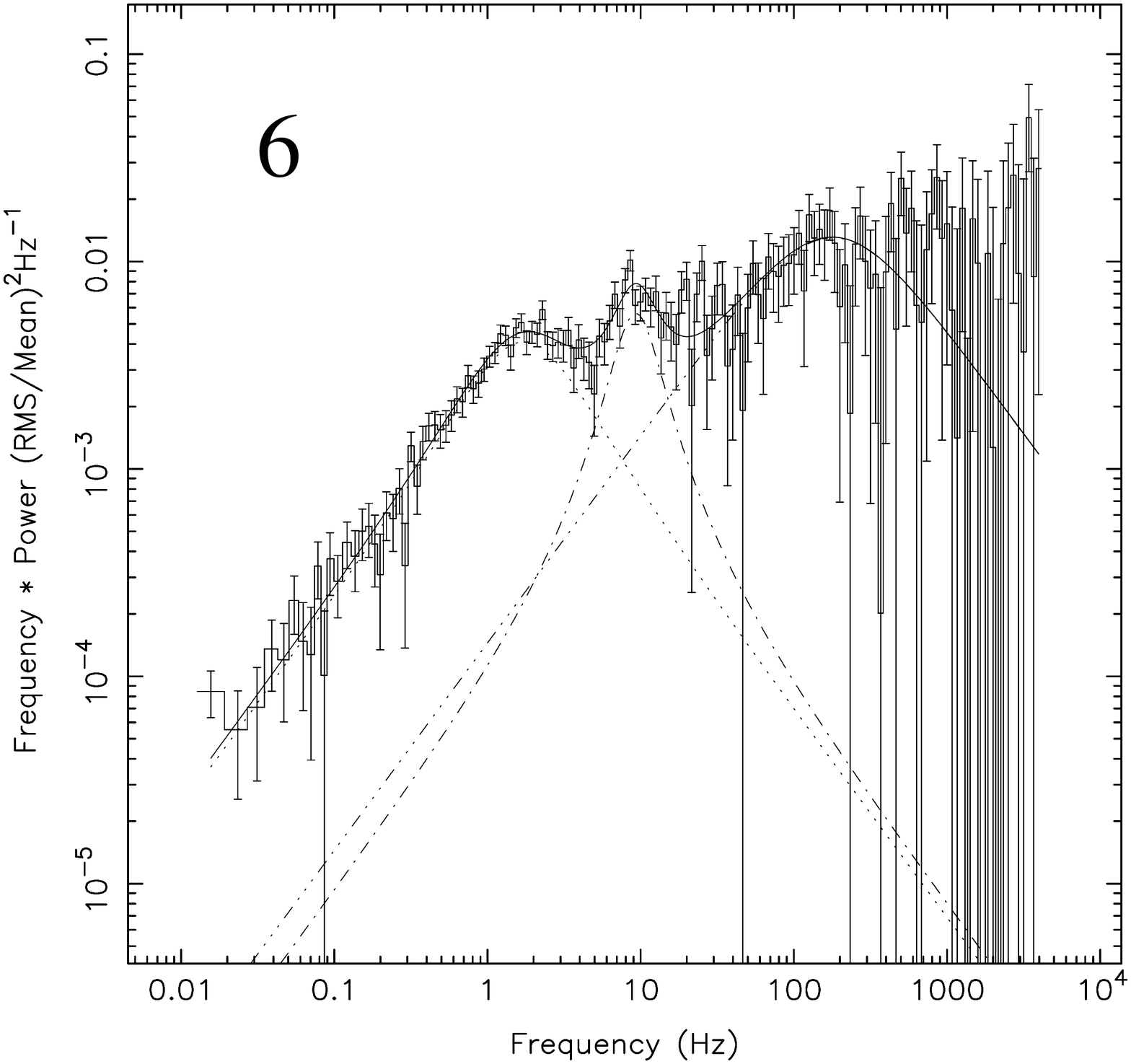}
\includegraphics[width=6.0cm,height=6cm]{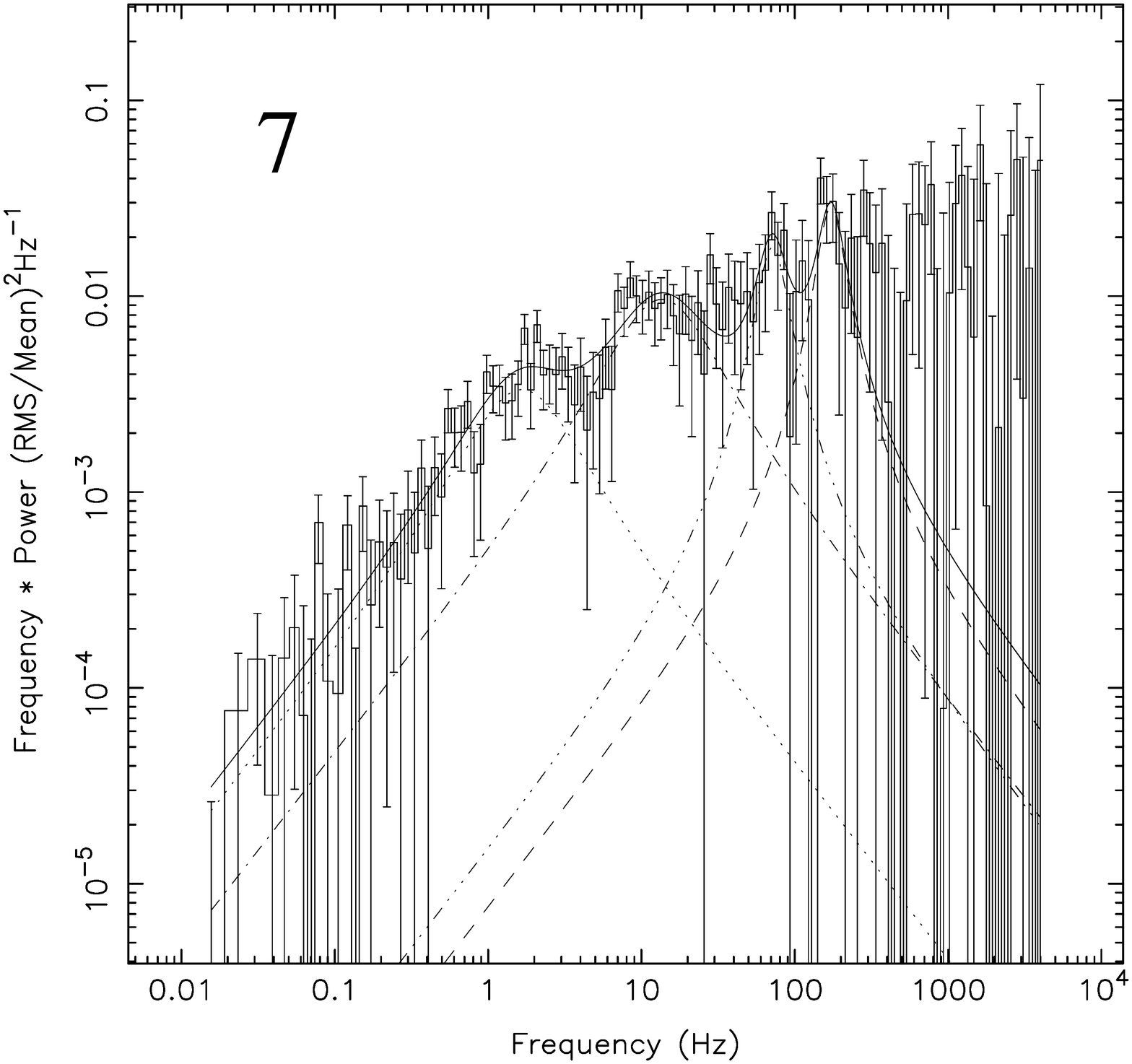}\includegraphics[width=6.0cm,height=6cm]{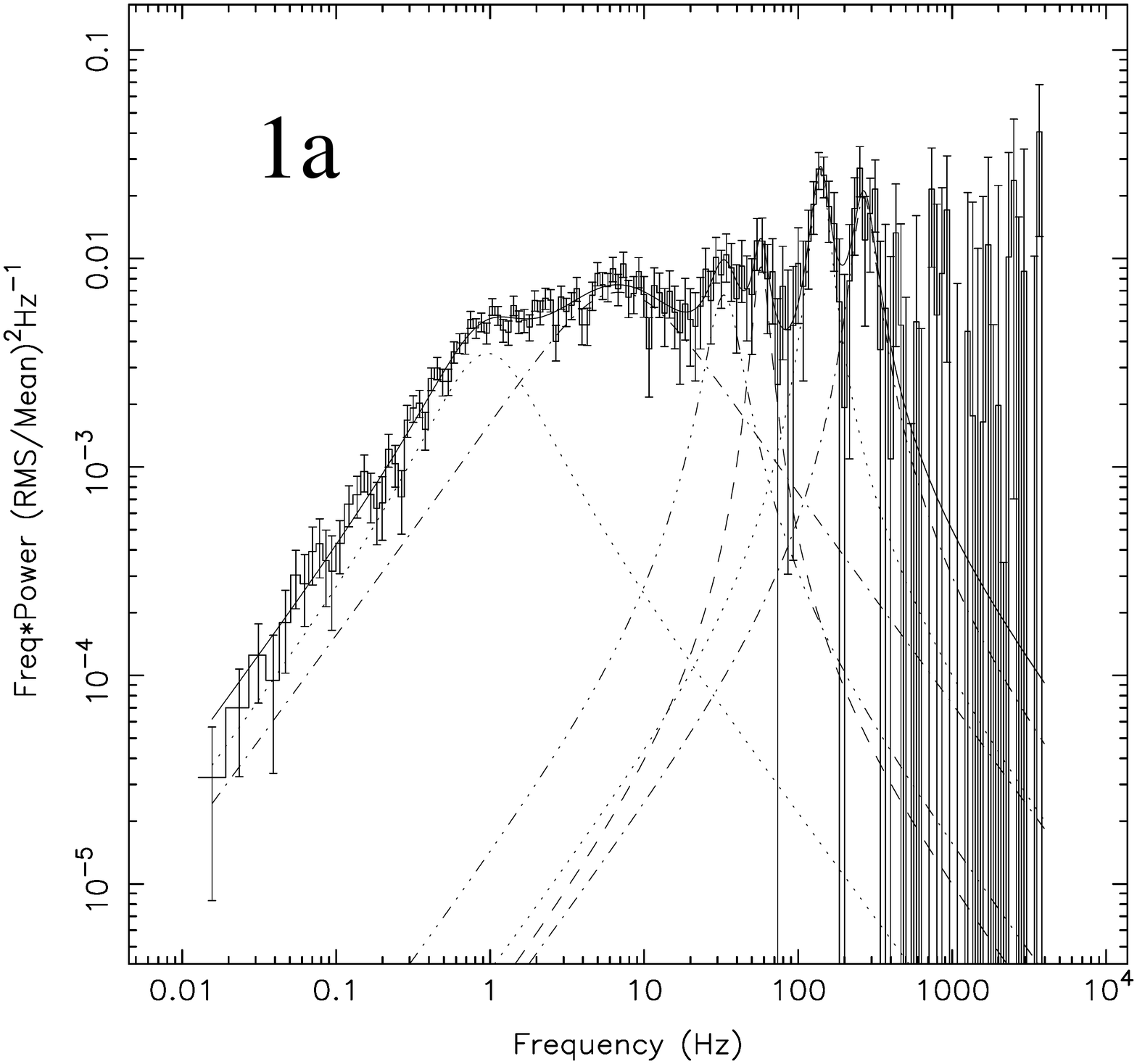}\includegraphics[width=6.0cm,height=6cm]{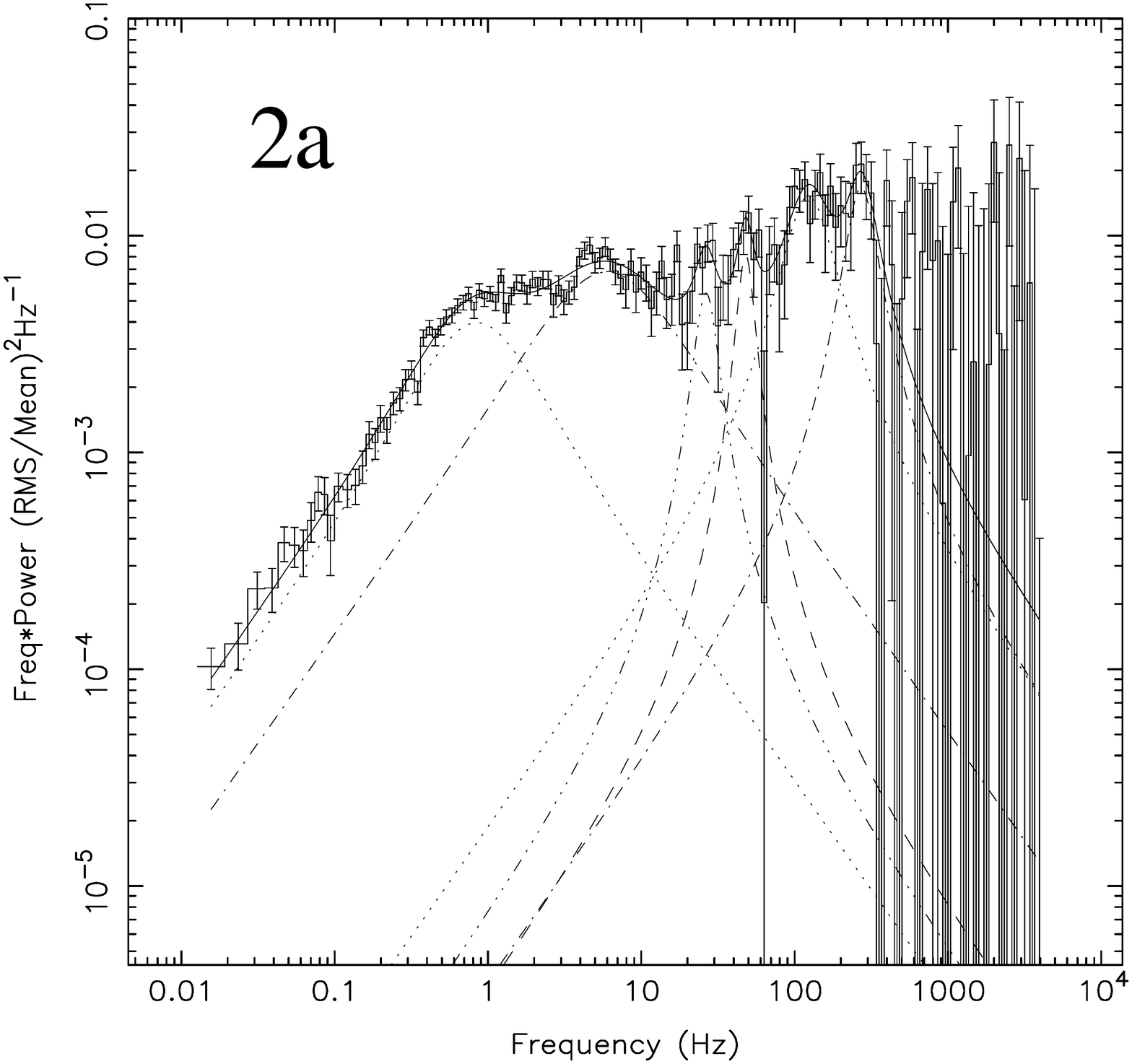}
\caption{Power spectra of the seven groups and groups 1a and 2a (using six Lorentzians instead of five as in groups 1 and 2, see Section \ref{sec-resultsap}) with multi-Lorentzian fit functions (see Table 1 for the best-fit parameters). Group numbers are indicated. The pulsar spike has been removed before rebinning in frequency. } \label{fig:pdsgrp}
\end{figure*}

The multi-Lorentzian fits to the power spectra of the seven groups are shown in Figure \ref{fig:pdsgrp}, with the best-fit parameter values listed in Table. 1. We report a conservative measure of the single trial significance ($\sigma$) of all parameters which is given by $P/P_-$, where P is the rms normalized power and $P_-$ is the negative error on P and is calculated using $\Delta\chi^2$ = 1. For a good fit (reduced $\chi^{2}$ $\sim$ 0.8$-$1.2, see Table 1) to the power spectra, 3--5 Lorentzians are needed. In some groups, the coherences of some components have been fixed \footnote{ When the broad noise components give a negative coherence, {\it{Q}} is fixed to 0 (which means it is a Lorentzian centered at zero frequency). For the other fixed parameters, the value of the fixed parameter is chosen to be the best-fit parameter value obtained when all the parameters are free.} as they are insufficiently constrained by the data. The fixing of these parameters does not affect the values of the other parameters. Groups 3--6 require three Lorentzians and show power spectra similar to each other. Interesting components at high frequencies are seen in groups 1, 2 and 7,  which therefore require additional Lorentzians in the fit. In each of groups 1 and 2, two high frequency QPOs ($\sim$135 and $\sim$260 Hz) are seen at similar characteristic frequencies. In group 7, it seems that the same two components appear but moved to much lower frequencies ($\sim$70 and $\sim$180 Hz); the centroid frequency difference $\Delta \nu$ between the two simultaneous peaks is similar to that seen in groups 1 and 2 between 100 and 150 Hz. \\
 \\
\begin{table}
\center
\renewcommand{\arraystretch}{0.65}
\begin{tabular}{cccccc}
\hline 
\hline
Grp. & $\nu_{max}$ (Hz) & Q & RMS (\%) &  $\sigma$  & Ident.\\
\hline
& 1.1 $\pm$ 0.1 & 0.3 $\pm$ 0.1  & 9.8 $\pm$ 1.1 & 4.2 & $L_{b}$ \\
& 6.4 $\pm$ 0.6 & 0.3 $\pm$ 0.3 & 11.7 $\pm$ 1.7 & 3.8 & $L_{h}$ \\
1 & 44.7 $\pm$ 7.1 & 0.5 $\pm$ 0.3 & 12.0 $\pm$ 1.8 & 4.1 & $L_{\ell ow}$\\
& 139.7 $\pm$ 4.2 & 3.3 $\pm$ 1.1 & 10.0 $\pm$ 1.4 & 3.8 & $L_{\ell}$ \\
& 251.8 $\pm$ 13.9 & 4.3 $\pm$ 2.8 & 9.3 $\pm$ 2.0 & 3.1 & $L_{u}$ \\
\hline
& 1.05 $\pm$ 0.1 & 0.25 $\pm$ 0.1 & 11.1 $\pm$ 0.7 & 12.3 & $L_{b}$ \\
& 5.2 $\pm$ 0.2 & 0.74 $\pm$ 0.3 & 9.3 $\pm$ 1.1 & 4.6 & $L_{h}$ \\
2 & 36.6 $\pm$ 7.5 & 0.16(fixed) & 13.2 $\pm$ 0.7 & 9.1 &  $L_{\ell ow}$ \\
& 129.9 $\pm$ 11.0 & 1.3(fixed) & 11.6 $\pm$ 1.3 & 3.7 &  $L_{\ell}$ \\
& 272.2 $\pm$ 13.9 & 2.45 $\pm$ 1.6 & 10.0 $\pm$ 2.5 & 3.3 & $L_{u}$ \\
\hline
& 1.3 $\pm$ 0.1 & 0.23 $\pm$ 0.1 & 10.2 $\pm$ 0.8 & 6.3 & $L_{b}$ \\
3 &7.4 $\pm$ 0.6 & 0.32 $\pm$ 0.2 & 11.2 $\pm$ 1.4 & 4.5 & $L_{h}$ \\
& 173.8 $\pm$ 16.9 & 0.23 $\pm$ 0.2 & 19.8 $\pm$ 1.1 & 9.1 & $L_{u}$ \\
\hline
& 1.65 $\pm$ 0.1 & 0.28 $\pm$ 0.1 & 10.8 $\pm$ 0.4 & 14.9 & $L_{b}$ \\
4 &7.4 $\pm$ 0.4 & 1.2 $\pm$ 0.3 & 7.5 $\pm$ 0.8 & 5.4 & $L_{h}$ \\
& 169.8 $\pm$ 22.4 & 0.0(fixed) & 21.2 $\pm$ 0.9 & 12.6 & $L_{u}$ \\
\hline
& 2.3 $\pm$ 0.2 & 0.1 $\pm$ 0.1 & 11.2 $\pm$ 0.4 & 14.6 & $L_{b}$ \\
5 & 8.3 $\pm$ 0.2 & 3.3 $\pm$ 0.8 & 6.1$\pm$ 0.6 & 5.4 & $L_{LF}$  \\
& 207.0 $\pm$ 27.0 & 0.0(fixed) & 23.5$\pm$ 1.0 & 11.9 & $L_{u}$ \\
\hline
& 1.7 $\pm$ 0.0 & 0.3 $\pm$ 0.1 & 9.75 $\pm$ 0.3 & 8.1 & $L_{b}$ \\
6 & 9.3 $\pm$ 0.5 & 1.2 $\pm$ 0.3 & 7.7 $\pm$ 0.7 & 5.8 & $L_{h}$  \\
& 180.0 $\pm$ 30.5 & 0.0(fixed) & 19.9 $\pm$ 0.9 & 11.3 & $L_{u}$ \\
\hline
& 1.64 $\pm$ 0.2 & 0.43 $\pm$ 0.1 & 8.3 $\pm$ 0.8 & 4.8 & $L_{b}$ \\
7 & 13.7 $\pm$ 2.5 & 0.43 $\pm$ 0.2  & 14.2 $\pm$ 1.5 & 5.5 & $L_{h}$ \\
& 72.5 $\pm$ 4.9 & 2.3(fixed) & 11.2 $\pm$ 1.4 & 4 & $L_{\ell}$ \\
& 179.9 $\pm$ 14.9 & 2.0(fixed) & 13.8 $\pm$ 1.5 & 4.5 & $L_{u}$ \\
\hline
& 0.93$\pm$0.1 & 0.51$\pm$0.2 & 7.93$\pm$1.2 & 3.0 & $L_{b}$ \\ 
& 6.85$\pm$0.8 & 0.06$\pm$0.2 & 14.1$\pm$1.2 & 6.0 & $L_{h}$ \\
1a & 30.9$\pm$1.2 & 4.1$\pm$2.8 & 5.4$\pm$1.1 & 2.7 & $L_{\ell ow/2}$ \\
& 56.9$\pm$1.8 & 5.1$\pm$3.8 & 6.2$\pm$1.0 & 3.2 & $L_{low}$ \\
& 140.3$\pm$3.4 & 3.95$\pm$1.2 & 10.34$\pm$1.1 & 4.9 & $L_{\ell}$\\
& 262.1$\pm$18.1 & 2.83$\pm$2.1 & 9.9$\pm$1.8 & 2.7 & $L_{u}$ \\
\hline 
& 0.86$\pm$0.1 & 0.3$\pm$0.1 & 9.7$\pm$0.9 & 4.7 & $L_{b}$ \\
& 5.95$\pm$0.4 & 0.2$\pm$0.2 & 13.3$\pm$1.5 & 4.2 & $L_{h}$ \\
2a & 26.4$\pm$1.3 & 2.6$\pm$1.7 & 5.5$\pm$2.1 & 2.4 & $L_{\ell ow/2}$ \\
& 47.9$\pm$1.9 & 3.6$\pm$1.9 & 5.8$\pm$1.1 & 2.4 & $L_{low}$ \\
& 123.0$\pm$8.6 & 1.2$\pm$0.4 & 12.8$\pm$1.4 & 5.5 & $L_{\ell}$ \\
& 271.6$\pm$14.5 & 2.0(fixed) & 10.9$\pm$1.2 & 4.4 & $L_{u}$ \\
\hline
\end{tabular}

\caption{}
\tablecomments{Best-fit parameters viz. Characteristic frequency $\nu_{max}$, Quality factor {\it{Q}} and Fractional rms. Also given are single-trial significance $\sigma$ of the Lorentzian feature defined as $P/P_-$, see Section \ref{sec-resultsap}, tentative component identification (Ident.) (based on scenario 1, see Section \ref{sec-highfreqcomident} ) of the 7 groups. Errors are at 1 standard deviation. } \label{table:para}
\end{table}
The average power spectrum of just the five brightest observations in the peak of the outburst (which are the last 5 observations of group 1, hereinafter group 1a) is shown in Figure \ref{fig:pdsgrp}. The 44.7 Hz component seen in group 1, in group 1a is resolved into two components with characteristic frequencies of 30.9 and 56.9 Hz. Similar behavior is exhibited by group 2 if we use six Lorentzians instead of five in the fit: the 36.6 Hz component is resolved into two components at 26.0 and 47.0 Hz. The best fit parameter values are given in Table 1. The centroid frequencies for group 1a are 30.9$\pm1.2$ and 56.6$\pm1.8$ Hz, and for group 2a (the six Lorentzian fit of group 2) they are 25.7$\pm1.2$ and 46.6$\pm1.0$ Hz. So, these two components are close to being harmonics of each other in both cases. The reason why they are not exactly harmonics could be that with time the components moved slightly in frequency, varying in strength in different ways. Averaging the power spectra can then mask an exact harmonic frequency relationship \citep{Mendez1998}. We note that although not all components in groups 1a and 2a have $P/P_-$ $>3$, they all appear in both, statistically independent, groups. 

\subsection{Identification of the components}
\label{sec-identification}
\subsubsection{Low-frequency Complex}
\label{sec-lowfreqcomident}
\begin{figure}
\centering
\includegraphics[width=8cm, height=6cm]{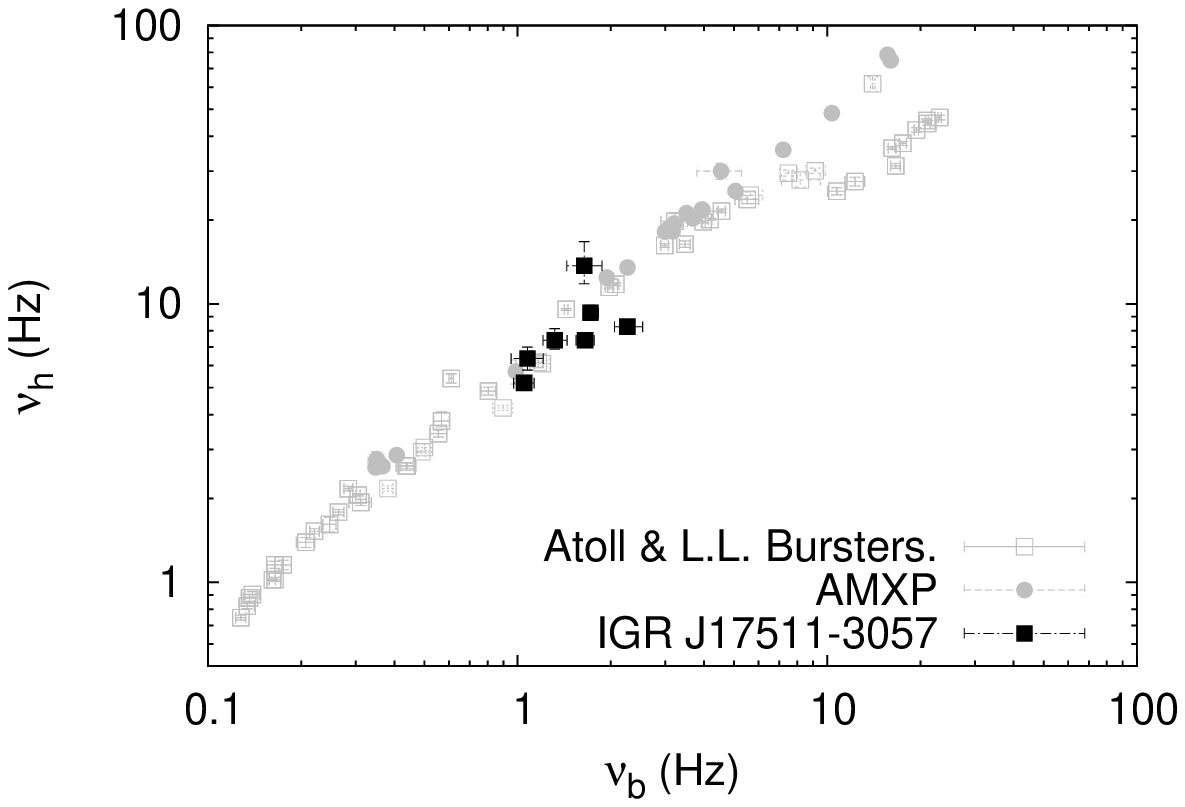}
\caption{Relation between $L_h$ and $L_b$ frequencies (WK relation, see Section \ref{sec-identification}) of the seven groups in IGR J17511-3057 compared to other atoll sources (grey points; from \citeauthor{2005} 2005).} \label{fig:hbrk}
\end{figure}

\begin{figure}
\center
\includegraphics[width=8cm, height=6cm]{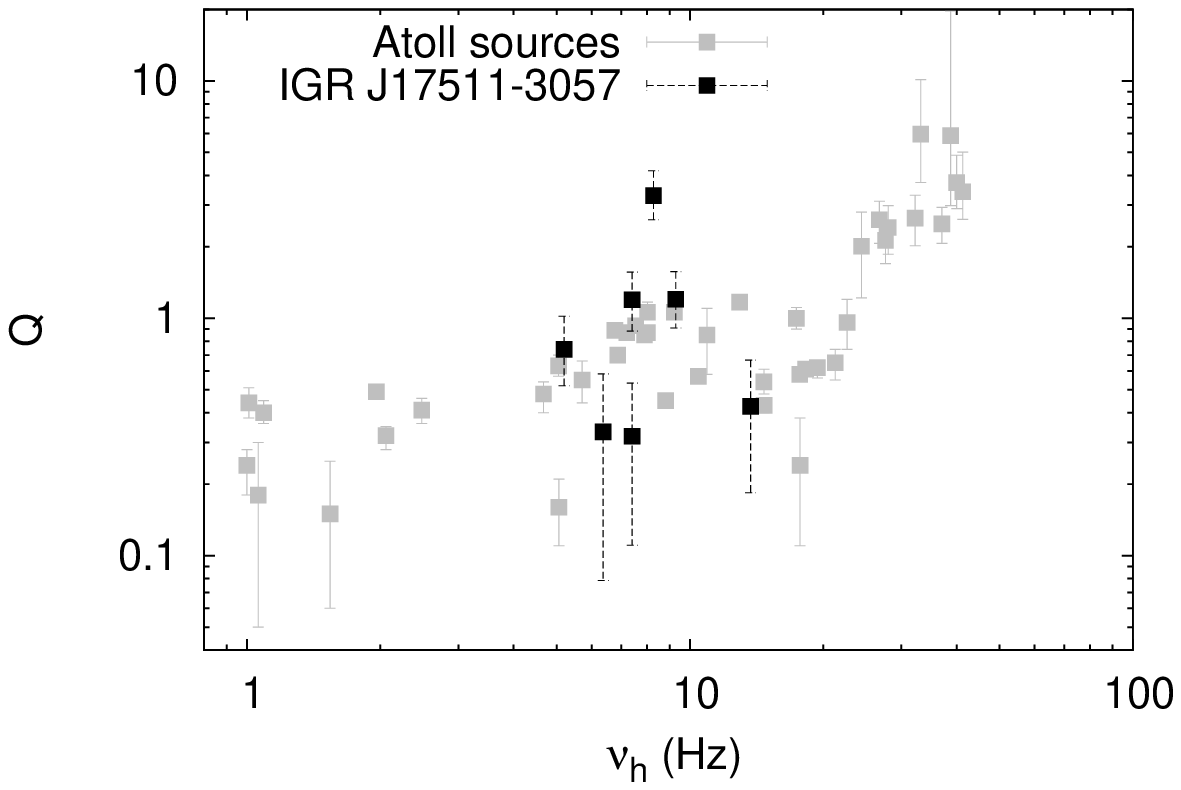}
\caption{Relation of coherence {\it{Q}} with characteristic frequency for $L_h$ in IGR J17511-3057 compared to other atoll sources (gray points; from \citeauthor{2005} 2005).} \label{fig:qhump}
\end{figure}

The average power spectra we observe at low frequencies resemble those seen in AMXPs  and other atoll sources in the EIS (see, e.g., \citeauthor{2006} 2006). These sources are known to exhibit correlations among the frequencies of the power spectral components (\citeauthor*{Wijnands1999} 1999, \citeauthor*{Psaltis1999} 1999, \citeauthor{2005} 2005). We use these correlations as a tool to attempt to identify the components in our source. Figure \ref{fig:hbrk} shows the relation between $L_b$ and $L_h$ frequencies as observed in many atoll sources known as the WK relation \citep{Wijnands1999}, along with the data points of our source. The two lowest frequencies in each group follow this correlation. So, they appear to be $L_{b}$ and $L_{h}$, respectively. Similarly, the correlation of the coherence Q with characteristic frequency is shown in Figure \ref{fig:qhump} for $L_h$ along with that in our source. The data for all but one groups fit the correlation. The feature at 8.3 Hz of group 5 deviates as it has a high coherence of 3.3. This feature could be a low frequency QPO ($L_{LF}$) as has been observed before in AMXPs \citep{2005} and other atoll sources \citep[see, e.g.,][for a review]{2006}. \\

\begin{figure}
\center
\includegraphics[width=11.5cm, height=10cm]{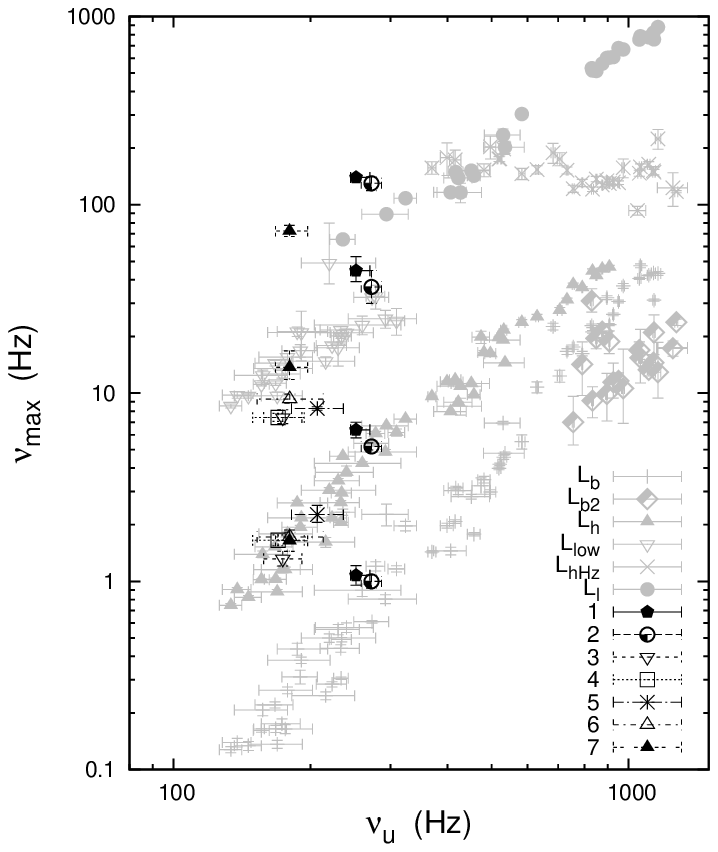}
\caption{ Characteristic frequencies of all the power spectral components plotted versus the characteristic frequency of $L_u$. The gray points are non-pulsating LMXBs. The data are from \cite{Altamirano2008}, and for Cir X-1, \cite{Boutloukos2006}. The seven groups of IGR J17511-3057 (black points) are indicated.} \label{fig:oldnunu}
\end{figure}

Figure \ref{fig:oldnunu} shows the existing the correlations between the characteristic frequencies of the different components in atoll sources versus the $L_u$ frequency. Assuming the highest frequency component in our source to be $L_u$, the data of the seven groups are plotted. From the match of groups 1 and 2 to other sources, the lowest frequency components appear to be $L_b$ and $L_h$, respectively. For the other groups, contrary to indications from Figure \ref{fig:hbrk}, the lowest frequency components are instead closer to the $L_h$ and $L_{\ell ow}$ tracks. The fractional rms amplitude of the individual low frequency components in all groups are similar to the observed values of the same components in other atoll sources in EIS \citep[see][for a comparison with AMXPs]{2005}. \\
\\
It is known that the correlations in AMXPs are often shifted relative to those in other atoll sources.  Shift factors upto 1.6 (see \citeauthor{2005} 2005, \citeauthor{Linares2005} 2005, \citeauthor{Linares2007} 2007, for details) have been observed so far. Except for groups 1 and 2 (where no shift factors are required for $L_b$ and $L_h$), we would require a shift factor 2.05 (obtained using the same method as \citeauthor{2005} 2005) to make our two lowest frequency components fall on the existing correlations if we interpret them as $L_b$ and $L_h$. As there is no single shift factor that would make all the data points of our source fit in the existing scheme of correlations, we do not apply any shifts in this paper, but consider other possibilities for this discrepancy between the correlations in Figures \ref{fig:hbrk} and \ref{fig:oldnunu}. We first discuss scenarios for the broad low frequency components:\\

\begin{figure}
\center
\includegraphics[width=8cm, height=6cm]{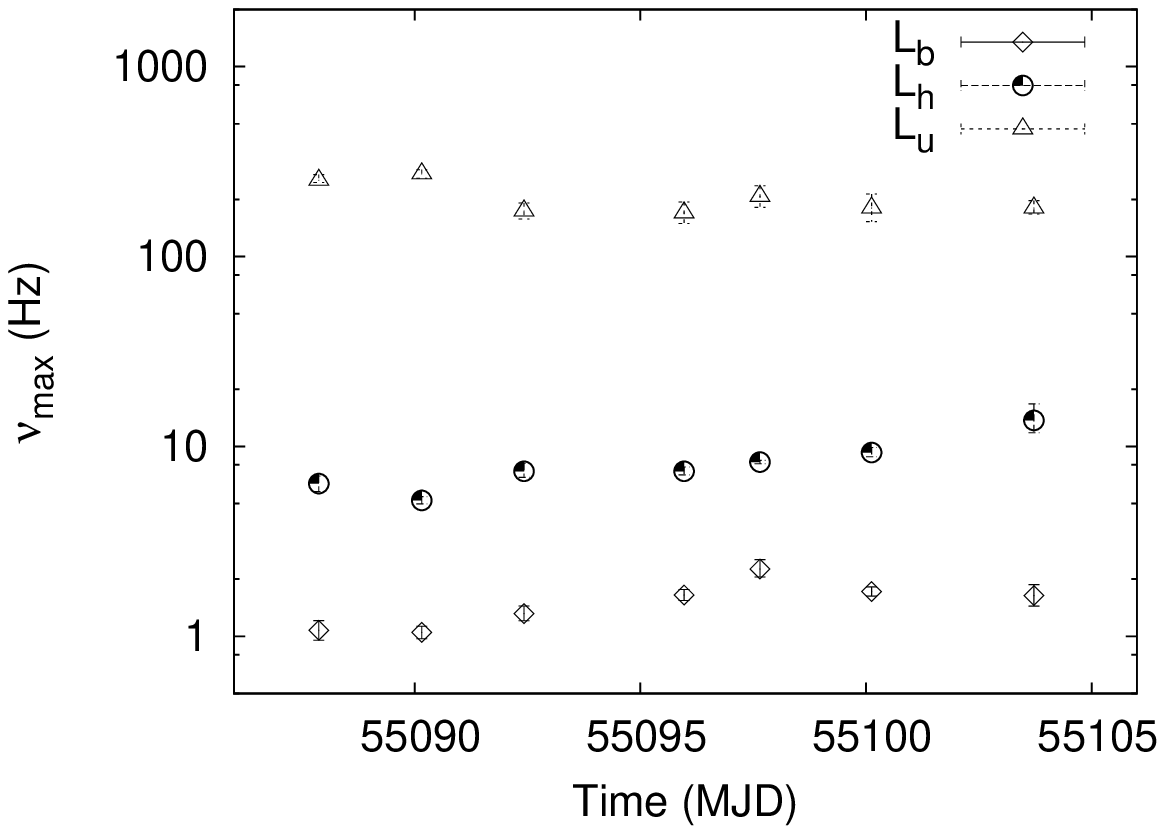}
\caption{Frequencies of the two lowest frequency components ($L_b$ and $L_h$) and the highest frequency components ($L_u$) of the seven groups in IGR J17511-3057 as a function of time.} \label{fig:frqvstime}
\end {figure}
[A]. The two lowest frequency components are $L_b$ and $L_h$ respectively. Figure \ref{fig:frqvstime} shows the plot of the frequency of the two lowest-frequency and the highest-frequency component in each group , identified as $L_b$, $L_h$ and $L_u$, respectively, as a function of time. These components are seen in all the power spectra and their frequencies do not change much in time, so it appears that we are seeing the same three components in all the power spectra. Yet only the $L_b$ and $L_h$ of groups 1 and 2 fall on the $L_b$ versus $L_u$ and $L_h$ versus $L_u$ relation. This might be because the $L_u$ from the fits of groups 3--6 are at lower frequencies than usual as only a broad, blended, component can be fit due to poor statistics, which does not allow us to resolve other components present, if any. However, such a single broad high frequency component is common in similar sources in the EIS, and these do show $L_b$ and $L_h$ on the standard relation. For group 7, a narrow $L_u$ is seen so in this case that explanation for the discrepancy does not apply. Together, this makes a reliable identification of $L_b$ and $L_h$ difficult for groups 3--7. \\

[B]. The two lowest frequency components are $L_b$ and $L_h$ in groups 1 and 2, and in the rest of the groups they are $L_h$ and $L_{\ell ow}$ respectively, as suggested by the correlations in Figure \ref{fig:oldnunu}. This does not support our conclusion of observing the same components in all the seven groups inferred from Figure \ref{fig:frqvstime}. \\
\subsubsection{High frequency QPOs}
\label{sec-highfreqcomident}
The power spectra of groups 1 and 2 are unusual, in that the frequencies of $L_b$ and $L_h$ are appropriate to the EIS, but presence of high-Q twin high frequency QPOs is more like what we would expect in the lower left banana (LLB) branch (without the very low frequency noise, VLFN). The fractional rms amplitudes of these high frequency components in groups 1 and 2 are similar to the observed rms values of the twin kHz QPO components in other atoll sources in the LLB. The second highest frequency QPOs of groups 1, 2 and 7 form a line well above the $L_{\ell ow}$ versus $L_u$ relation \footnote{It has been suggested previously by \cite{2005} that $L_{\ell}$ and $L_{{\ell}ow}$ are different components} as seen in Figure \ref{fig:oldnunu}. It is interesting to note that this line is also above the $L_\ell$ versus $L_u$ relation of other atoll sources. If we use the factor of 2.05 derived earlier in Section \ref{sec-lowfreqcomident} to shift the high frequency components in $\nu_u$ and $\nu_\ell$, they do fall on the existing $\nu_\ell$ versus $\nu_u$ correlation, but then the corresponding low frequency components do not fall on their respective correlations (for works done earlier see \citeauthor{2005} 2005 and \citeauthor{Linares2005} 2005). Hence, as discussed in Section. \ref{sec-lowfreqcomident}, we do not rely on the shift factors to help us identify the high frequency components. There are a number of possibilities for the identification of the high frequency components in groups 1, 2 and 7. We discuss below the possible scenarios:\\

[1]. The high frequency QPOs seen in groups 1, 2 and 7 are the twin kHz QPOs. The components at 44.7 Hz in group 1 and at 36.6 Hz in group 2 are $L_{\ell ow}$ as they fall close to the $L_{\ell ow}$ versus $L_u$ correlation. $L_{\ell ow}$ is expected to be seen in the EIS. However, harmonics of $L_{\ell ow}$ have not been observed before, nor have twin kHz QPOs been seen in the EIS (with one exception, 4U 1728-34; \citeauthor{Migliari2003} 2003). \\

[2]. The high frequency QPOs seen in groups 1, 2 and 7 are the twin kHz QPOs. The components at 44.7 Hz in group 1 and at 36.6 Hz in group 2 are $L_{hHz}$. kHz QPOs are often accompanied by $L_{hHz}$ components and harmonics of $L_{hHz}$ were observed once previously in 4U 1636-53 \citep{Altamirano2008}. However, the putative $L_{hHz}$ components have frequencies that are too low compared to all other sources, and as mentioned above, twin kHz QPOs are not expected in the EIS. \\

[3]. The highest frequency QPOs are $L_u$ and the second highest frequency QPOs are $L_{hHz}$. The components at 44.7 Hz and 36.6 Hz are $L_{ \ell ow}$ as in scenarion 1. The $2^{nd}$ highest frequency QPO in group 7 could be either $L_{hHz}$ or $L_{\ell ow}$; this is not clear from the correlation. The drawback of this interpretation is that $L_{hHz}$ and $L_{\ell ow}$ have never been observed together. $L_{hHz}$ is not expected in the EIS as observed from previous works: as seen in Figure \ref{fig:oldnunu}, there are no $L_{hHz}$ components observed at low $L_u$ values characteristic for the EIS \citep[see, e.g.,][]{2006}. \\

[4]. The two highest frequency QPOs in groups 1 and 2 are $L_{hHz}$ features. At centroid frequencies $138.1^{+4.4}_{-4.1}$, $250.1^{+18.6}_{-7.3}$ and $121.8^{+10.2}_{-10.4}$, $266.7^{+15.6}_{-15.0}$,  respectively ($139.2\pm3.4$, $258.1\pm18.7$ and $113.5\pm9.7$, $263.5\pm14.0$ for groups 1a and 2a), they are all approximate harmonics of each other and as discussed in Section \ref{sec-resultsap}, the features might have moved in time masking an exact harmonic relation. As $L_u$ is not present in this scenario, we cannot use the correlations in Figure \ref{fig:oldnunu} to identify the other components. The components at 44.7 Hz and 36.6 Hz are $L_{\ell ow}$ as in scenario 1. The drawbacks of this scenario are that the $L_{hHz}$ features are not expected to be observed in the EIS and the $L_{hHz}$ and $L_{\ell ow}$ have not been observed together. Also, the two highest frequencies of group 7 cannot be identified in this scenario, as they are not harmonically related. \\
\\
It is clear that every scenario has its own shortcomings, which makes a reliable identification difficult. If we try to identify the components just by comparing the power spectra with those of other sources, the identification does not comply with the state as deduced from the CD pattern. The source has some unusual high frequency components seen in groups 1, 2 and 7 and an unusual triplet of broad low frequency components in groups 3--6. The possibility that the highest frequency feature is not $L_u$ cannot be ruled out, especially for group 7. No single shift factor such as proposed for other AMXPs can restore a full match to the other atoll sources.
\begin{figure}
\center
\includegraphics[width = 8.5cm]{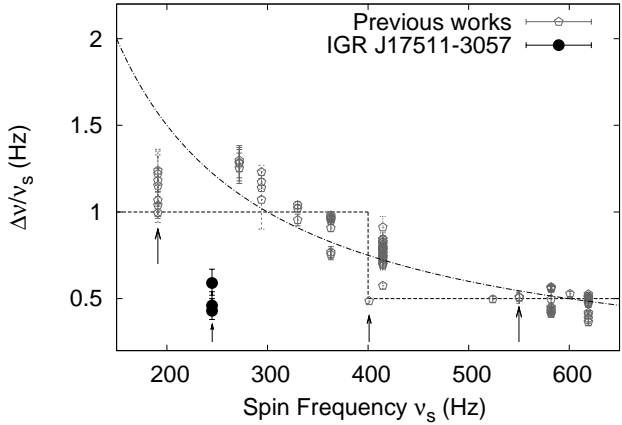}
\caption{Ratio of measurements of $\Delta\nu$ = $\nu_u$ -- $\nu_\ell$ and spin frequency $\nu_s$ as a function of $\nu_s$ in IGR J17511--3057 and other NS LMXBs (gray points; \citeauthor{Altamirano2010b} 2010b). The step function is $\Delta\nu$/$\nu_s$ = 1 for $\nu_s\leq$ 400 Hz and $\Delta\nu$/$\nu_s$ = 0.5 for $\nu_s\geq$ 400 Hz. The curved line corresponds to $\Delta\nu$ = 300 Hz. The four AMXPs are marked with arrows.} \label{fig:deltanuvsnu}
\end{figure}
\section{Discussion}  
\label{sec-discussion}
The behavior of IGR J17511--3057 in the CD and the power spectra at first sight appears similar to what has been observed in other atoll sources and AMXPs \citep[see, e.g,][]{2006}. From the color diagrams and the shape of power spectra of groups 3--6, the source appears to be an atoll source in the EIS. The other AMXPs have also been classified as atoll sources (see \citeauthor{Wijnands2006} et al. 2006, \citeauthor{Watts2009b} 2009a, \citeauthor{Linares2008} 2008, \citeauthor{Kaaret2003} 2003, \citeauthor{Reig2000} 2000). However, closer study of the power spectral components indicates that this source is peculiar and does not fit well in the scheme defined by other sources. As discussed in Section \ref{sec-identification}, all the possible scenarios for fitting our source in this scheme have their own shortcomings. Therefore, none of the components can be identified with certainty.\\
\\
The scenario that appears to require the least number of additional assumptions is scenario 1, where the only peculiarity is that twin kHz QPOs appear in what is otherwise an ordinary EIS; this was reported once before, in 4U 1728--34 \citep{Migliari2003}. We note that if the two high frequency components are twin kHz QPOs, the measured differences $\Delta\nu$ between the centroid frequencies of high frequency QPOs in groups 1, 2 and 7 are $112^{+19}_{-8}$ , $144.9^{+18.6}_{-18.2}$ and $104.2^{+17.2}_{-12.9}$ Hz, respectively, and $119^{+22}_{-15.6}$ and $150^{+18.1}_{-15.8}$ Hz in group 1a and group 2a, respectively. These values are inconsistent with being close to the spin frequency $\nu_{s}$ of 244.8 Hz  as would have been expected for this so-called slow rotator ($\nu_{s}$ $<$ 400 Hz; \citeauthor{Miller1998} 1998). Instead, they are all consistent with half the spin frequency, which otherwise has only been seen in fast rotators ($\nu_{s}$ $>$ 400 Hz). This can be seen in Figure \ref{fig:deltanuvsnu} which shows the plot of $\Delta\nu$/$\nu_s$ as a function of $\nu_s$ for AMXPs and other atoll sources (the spin frequency is inferred from burst oscillations for these systems). The step function shows the historical distinction between the slow and fast rotators. It has been suggested that $\Delta\nu$ and $\nu_s$ are (nearly) independent (\citeauthor{Yin2007} 2007, \citeauthor{M'endez2007} 2007). The curved line represents a constant $\Delta\nu$ of 300 Hz. To make the AMXPs SAX J1808.4--3658 and XTE J1807--294 fit this curve, they would have to be shifted up by a factor of $\sim$1.5 \citep{M'endez2007}. The points of IGR J17511--3057 for groups 1, 2 and 7 would require a factor $\sim$2.5 to fall on this curved line. However, groups 1 and 2 do not require this same factor to fit in the scheme of correlations seen in Figure \ref{fig:oldnunu}, but rather a factor 2.05. We applied no shifts to any data in this paper as there is no single factor. Note that all four AMXPs in Figure \ref{fig:deltanuvsnu} are consistent with either $\Delta\nu$ = $\nu_s$ or $\Delta\nu$ = $\nu_{s/2}$ without shifts. \\
\\
Our results favor models like the sonic-point and spin-resonance model \citep{Lamb2003} and the relativistic resonance model \citep{Kluzniak} which predict that either $\Delta\nu$ = $\nu_s$ or $\Delta\nu$ = $\nu_s$/2. The relativistic precession model \citep{Stella1999} predicts that $\Delta\nu$ should decrease when $\nu_u$ increases as well as decreases (see Figure 2.14. in \citeauthor{2006} 2006). Our results do suggest a low $\Delta\nu$, however the value is almost a factor of two lower than the value expected from the model at the observed $\nu_u$ of IGR J17511--3057. \\

In conclusion, IGR J17511--3057 is indeed a very peculiar and interesting source. If scenarios 1 and 2 apply, the properties of the source can be summarized as follows: 
a) It exhibits kHz QPOs while in the EIS,
b) In spite of being a slow rotator, kHz QPO frequency separation is $\Delta\nu$ $\sim$ $\nu_s$/2 and
c) It requires different shift factors to fall on the frequency correlations of LMXBs at different times.
Clearly this source could play a very important role in testing the existing models for the origin of QPOs. More observations of this source with {\it{RXTE}} or perhaps ASTROSAT \citep{Agrawal2002}, when it goes into an outburst again, are necessary to understand the nature of the different components. If we could observe the source in different spectral states, the components exhibited and their frequency evolution would help in establishing their nature. Observations of the high frequency QPOs and their evolution as a function of time and spectral state are key to their reliable identification.

\end{document}